\documentstyle[epsfig,12pt]{article}

\newcommand{\beq}{\begin{equation}}
\newcommand{\eeq}{\end{equation}}
\newcommand{\bea}{\vspace{0.25cm}\begin{eqnarray}}
\newcommand{\eea}{\end{eqnarray}}

\newcommand{\r}{\mbox{{\boldmath
$\rho$}}}

\newcommand{\qb}{\mbox{{\bf
q}}}
\newcommand{\pb}{\mbox{{\bf
p}}}

\newcommand{\qbt}{\mbox{{\bf
q}}_\perp}

\newcommand{\fb}{\mbox{{\bf
f}}}

\def\lsim{\mathrel{\rlap{\lower4pt\hbox{\hskip1pt$\sim$}}
    \raise1pt\hbox{$<$}}}         
\def\gsim{\mathrel{\rlap{\lower4pt\hbox{\hskip1pt$\sim$}}
    \raise1pt\hbox{$>$}}}         
\begin{document}
\thispagestyle{empty}
\vspace*{-2cm}
 
\bigskip
 
\begin{center}

  {\large\bf
PHOTON EMISSION FROM 
BARE QUARK STARS
\\
\vspace{1.5cm}
  }
\medskip
  {\large
B.G. Zakharov
}
  \bigskip

{\it
L.D. Landau Institute for Theoretical Physics,
        GSP-1, 117940,\\ Kosygina Str. 2, 117334 Moscow, Russia
\vspace{2.7cm}\\}
  {\bf
  Abstract}
\end{center}
{
\baselineskip=9pt
We investigate the photon emission from the electrosphere
of a quark star. It is shown that at temperatures $T\sim 0.1\div 1$ MeV
the dominating mechanism is the bremsstrahlung due 
to bending of electron trajectories in the mean Coulomb field of the
electrosphere. The radiated energy for this mechanism 
is much larger than that for the Bethe-Heitler 
bremsstrahlung.  The energy flux from the mean field bremsstrahlung 
exceeds the one from the tunnel $e^{+}e^{-}$ pair creation as well.
We demonstrate that the LPM suppression of the photon emission
is negligible.\\
}
\vspace{1cm}

\newpage

\section{Introduction}

The hypothesis of quark stars made of a stable strange quark 
matter (SQM) \cite{Witten,Alcock1,Haensel} has been attracting 
much attention for many years. 
It is possible that quark stars (if they exist) may be (at least in the 
initial hot stage) without a crust of normal matter \cite{Usov1}.
Contrary to neutron stars the density of SQM for bare quark stars should 
drop abruptly at the scale $\sim 1$ fm. 
The SQM in normal phase and in the two-flavor superconducting  (2SC) phase 
should also contain electrons 
(for normal phase the electron chemical potential, $\mu$, 
is about 
20 MeV \cite{Alcock1,Kettner}). 
In contrast to the quark density the 
electron density  drops smoothly above the star surface at the 
scale $\sim 10^{3}$ fm \cite{Alcock1,Kettner}. 
For the star surface temperature 
$T\ll \mu$, say $T\lsim 10^{10} \mbox{K}\sim 1$ MeV, 
this ``electron atmosphere'' (usually called the electrosphere) 
may be viewed as 
a strongly degenerate relativistic
electron gas \cite{Alcock1,Kettner}.

From the point of view of distinguishing bare quark stars from neutron stars 
it is of great importance to have theoretical predictions for
the photon emission from bare quark stars.
Contrary to neutron stars 
(or quark stars with a crust of normal matter)
the photon emission from quark stars made of a stable self-bound SQM
may potentially exceed the Eddington limit. This fact may be used for 
detecting a bare quark star.
However, the SQM itself is a very poor emitter 
at $T\ll \omega_{p}^{q}$ \cite{Chmaj,Harko1} (here 
$\omega_{p}^{q}\sim 20$ MeV  is the plasma frequency of the SQM
\cite{Chmaj}). At such temperatures the photon emission 
from the quark surface is a tunnel process, and the radiation rate turns out
to be negligibly small as compared
to the black body radiation \cite{Chmaj}.
However, for the electrosphere the plasma frequency, $\omega_{p}^{e}$, 
is much smaller
than that for the SQM. For this reason the photon emission from 
the electrosphere may potentially dominate the luminosity of a quark 
star. 
For understanding the prospect of detecting bare quark stars it is highly 
desirable to have 
quantitative predictions for the photon emission from the electrosphere.
This is also of interest in the context of the scenario of the gamma-ray 
repeaters due to reheating of a quark star by impact of a massive comet-like
object \cite{Usov-GRB}, and the dark matter model in the form of 
matter/antimatter SQM nuggets \cite{Zh2}.

An obvious candidate for the photon emission from the electrosphere 
is the bremsstrahlung from electrons. It may be due to either 
the electron-electron interaction (the Bethe-Heitler bremsstrahlung) 
or interaction of electrons with the 
mean electric field of the electrosphere.
One more mechanism is related to the 
tunnel $e^{+}e^{-}$ pair creation 
\cite{Usov1,Usov2}. The point is that the electric field of the electrosphere 
should be very strong. It may be about several tens of the critical 
field for the tunnel Schwinger pair production $E_{cr}=m_{e}^{2}/e$ 
\cite{Sch} (we use units $c=\hbar=k_{B}=1$).
In this scenario the photons appear through  $e^{+}e^{-}$ annihilation 
in the outflowing $e^{\pm}$ wind
\cite{Aks}.

The bremsstrahlung from the electrosphere due to the 
electron-electron interaction has been addressed
in \cite{Gale,Zh1}. 
The authors of \cite{Gale} used the soft photon approximation
and factorized the $e^{-}e^{-}\to e^{-}e^{-}$ cross section
in the spirit of Low's theorem.
In \cite{Zh1} it was pointed out that this approximation is inadequate
since it neglects the effect of the photon energy on the electron 
Pauli-blocking which should lead to a strong suppression of the 
radiation rate. The authors of \cite{Zh1} have not given a consistent
treatment of this problem either. To take into account the effect of the
minimal photon energy they suggested some restrictions on the initial electron
momenta imposed by hand.
In this way they obtained the 
radiated energy flux 
from the
$e^{-}e^{-}\to e^{-}e^{-}\gamma$ process
which is much smaller than 
that in \cite{Gale}, and than the energy flux
from the tunnel $e^{+}e^{-}$ pair creation \cite{Usov1,Usov2}. 
In \cite{Harko2} there was an initial 
attempt to include the effect of the mean Coulomb
field of the electrosphere on the photon emission. The authors obtained
a considerable enhancement of the radiation rate.
However, similarly to \cite{Gale} the analysis
\cite{Harko2} treats incorrectly the Pauli-blocking effect.
Note also that in the analyses \cite{Zh1,Harko2}
the photon quasiparticle mass was neglected. 
As we will show this approximation is clearly inadequate since the 
finite photon mass suppresses the radiation rate strongly.

Thus the theoretical situation with 
the photon bremsstrahlung from the electrosphere is still controversial 
and uncertain. The main problem here, which was not solved in the previous 
analyses \cite{Gale,Zh1,Harko2}, is an accurate accounting for 
the photon energy in the electron Pauli-blocking.
In the present paper we address the bremsstrahlung
from the electrosphere in a way similar to the Arnold-Moore-Yaffe
(AMY) \cite{AMY1}  approach to the collinear photon emission from 
a hot quark-gluon plasma based on the thermal field theory.
We use a reformulation of the AMY formalism given in \cite{AZ}.
It is based on the light-cone path integral (LCPI) 
approach \cite{Z1,Z_SLAC1,Z_SLAC2} (for reviews, see 
\cite{Z_YAF98,Z_NP2005}) to the in-medium radiation processes. 
For an infinite homogeneous plasma 
(with zero mean field) 
the formalism  \cite{AZ} reproduces the 
AMY results \cite{AMY1}.
The LCPI formulation \cite{AZ} has the advantage that it also works for 
plasmas with nonzero mean field. 
It allows to evaluate the photon emission accounting for 
bending of the electron trajectories in the mean Coulomb potential 
of the electrosphere.
Contrary to very crude and qualitative methods of
\cite{Gale,Zh1,Harko2} the treatment of the Pauli-blocking
effects in \cite{AMY1,AZ} has robust quantum field theoretical
grounds. 
Of course, our approach is only valid in the regime of collinear 
photon emission when the dominating photon energies exceed several 
units of the photon quasiparticle mass. Numerical calculations show 
that even at $T\sim 0.1$ MeV the effect of the noncollinear 
configurations is relatively small.  

We demonstrate that for the temperatures $T\sim 0.1\div 1$ MeV 
the radiated energy flux from the  $e^{-}\to e^{-}\gamma$ transition in 
the mean electric field  
turns out to be much larger than that from the
Bethe-Heitler bremsstrahlung.
It exceeds the energy flux from the tunnel $e^{+}e^{-}$ pairs as well. 
Also, we demonstrate that, contrary
to conclusion of \cite{Gale}, the Landau-Pomeranchuk-Migdal (LPM) suppression
\cite{LP,M} of the photon bremsstrahlung is negligible.
Our results show that the photon emission
from the electrosphere may be of the same order as the black body radiation. 
For this reason the situation with distinguishing a bare quark star 
made of the SQM in normal (or 2SC) phase from a neutron star
using the luminosity \cite{Usov1,Usov-LC} 
may be more optimistic
than in the scenario with the tunnel $e^{+}e^{-}$ pair creation \cite{Usov1}.

In a short version the results of this work were presented in \cite{Z_qs1}.
In this paper we present our results in a more detailed form.
The plan of the paper is as follows. In Sec.~2 we review the 
basic formulas and approximations.
In Sec.~3 we discuss the evaluation of photon emission from a given
electron in the electromagnetic field of the electrosphere which
includes both the mean Coulomb field and the ordinary fluctuation field
generated by neighboring electrons.
In Sec.~4  we present   
 numerical results for  the radiated energy flux.
Sec.~5 is devoted to conclusions.

\section{Basic  formulas and approximations}
As in Refs. \cite{Usov1,Gale,Zh1} we use for the electrosphere the model of 
a relativistic strongly degenerate electron gas in the Thomas-Fermi
approximation. In this approximation the local electron number density 
reads $n_{e}(h)=\mu^{3}(h)/3\pi^{2}$, where $h$ is the distance from the 
quark surface. The $h$ dependence of the chemical potential is governed
by the Poisson equation for the electrostatic potential 
$V=\mu/e$. For $h>0$ this gives 
\cite{Alcock1,Kettner} 
\beq
\mu(h)=\frac{\mu(0)}{(1+h/H)}\,,
\label{eq:10}
\eeq
where
$H=\sqrt{{3\pi}/{2\alpha}}/\mu(0)$, 
$\alpha=e^{2}/4\pi$.

We assume that the electrosphere is optically thin. 
This means that the photon absorption and stimulated emission can be neglected.
In this regime the luminosity
may be expressed in terms of the energy  radiated
spontaneously per unit time and volume, $Q$,
usually called the emissitivity.
In the formalism \cite{AZ} the emissitivity per unit photon energy $\omega$ 
at a given $h$  can be written as 
\beq
\frac{dQ(h,\omega)}{d\omega}=
\frac{\omega(k)}{4\pi^{3}}
\frac{dk}{d\omega} 
\int \frac{d\pb}{p} n_{F}(E)[1-n_{F}(E')]\theta(p-k)
\frac{dP(\pb,x)}{dx dL}\,,
\label{eq:20}
\eeq
where 
$k$ denotes the photon momentum, 
$E$ and $E'$ are the electron energies 
before and after photon emission, 
$n_{F}(E)=[\exp((E-\mu)/T)+1]^{-1}$ is the local electron Fermi
distribution (we omit the argument $h$ in the functions 
on the right-hand side 
of (\ref{eq:20})), and
$x=k/p$ is the photon longitudinal (along the initial electron momentum
$\pb$) fractional  
momentum. The function $dP/dx dL$ in (\ref{eq:20}) is the probability  
of the photon emission per unit $x$ and length from an electron in 
the potential generated by other electrons which includes both the 
smooth collective Coulomb field
and the usual fluctuating plasma part related to the field
generated by the neighboring electrons.
Note that the formula (\ref{eq:20}) accounts for photons 
emitted to all
directions, because in the optically 
thin electrosphere  practically all the photons 
radiated to the hemisphere directed to the quark surface will be reflected
either in the electrosphere (at the level with $\omega_{p}^{e}=\omega$)
or from the quark surface. Only the photons
with $\omega\gsim\omega_{p}^{q}\sim 20$ MeV may be absorbed in the quark
matter. However, such photons are not important at $T\lsim 1$ MeV 
considered in the present paper. For the above reasons, it would be incorrect
to exclude the photons emitted towards the star surface, as was done
in \cite{Zh1}.

Our basic formula (\ref{eq:20}) assumes that 
the photon emission is a local process, {\it i.e.} the photon 
formation length \footnote{Physically, the photon formation length (sometimes 
called the coherence length) is a longitudinal scale at which the photon 
and electron wave 
packets become separated. It appears naturally in the LCPI approach
\cite{Z1,Z_YAF98} formulated in the coordinate space as a dominating scale of
the integrals in the longitudinal coordinate.} (we denote it $l_{f}$) 
is small compared to the thickness of the 
electrosphere. 
Evidently, only in this case one can define a 
local emissitivity. Note that Eq. (\ref{eq:20}) defines the rate of photon
production at a given photon energy, which remains constant during the
photon propagation in the electrosphere. The photon momentum 
in this process changes adiabatically according to the photon quasiparticle 
dispersion relation in the electron plasma. 
Also, the formula (\ref{eq:20}) assumes
that on the scale $\sim l_{f}$ the electron trajectories are smooth.
It means that besides the evident condition $l_{f}\ll R_{m}$
($R_{m}$ is the radius of curvature of the electron trajectory in the mean 
field) the typical scattering angle related to the random walk of an 
electron due to electron-electron interaction should be small as well.
One can show that these conditions are satisfied for the 
electrosphere.
An important consequence of the smoothness of electron trajectories
at the scale $\sim l_{f}$ is the longitudinal factorization of the 
Pauli-blocking factor $[1-n_{F}(E')]$ for the final state of the 
radiating electron in (\ref{eq:20}). Namely the fact that the trajectories
are smooth in the process of the photon emission allows one to neglect
the statistics effects in treating the small angle scattering.
Indeed, the typical space scale for the soft fluctuating modes 
of the electromagnetic field is about the inverse 
Debye mass $1/m_{D}\sim 1/e\mu$. This scale
is much larger than the typical separation between electrons
$\sim 1/\mu$. For this reason from the point of view of the 
electrons with energy $\sim \mu$ the soft electromagnetic 
field at the space scale $\sim 1/m_{D}\gg 1/\mu$ can be viewed as a 
uniform field
at the scale $\sim 1/\mu$. In a uniform field all electrons 
in the same spin state scatter in the same way, and the small angle scattering
leads simply to some shift of the distribution function in the 
momentum space. Any statistics effects will be suppressed by some power of 
electron charge $e$.
Calculations  
within the real time thermal field theory 
performed in \cite{AMY1} corroborate this physical picture
of the collinear photon emission.

In our approximation of optically thin medium the differential
radiated energy flux from the electrosphere,
$dF/d\omega$, in terms of the emissitivity 
reads
\beq
\frac{dF}{d\omega}=\int_{0}^{h_{max}}dh 
\frac{d Q(h,\omega)}{d\omega}\,.
\label{eq:21}
\eeq
For the chemical potential (\ref{eq:10}) the $h$-integration
in (\ref{eq:21}) can be approximated by the integration over $\mu$ as
\beq
\frac{dF}{d\omega} 
\approx
\sqrt{\frac{3\pi}{2\alpha}}
\int_{\mu_{min}}^{\mu(0)}\frac{d\mu}{\mu^{2}}
\frac{dQ(h(\mu),\omega)}{d\omega}
\label{eq:22}
\eeq
with $\mu_{min}=\mu(h_{max})$.
In numerical calculations we take 
$\mu_{min}=2m_{e}$. Of course, the relativistic approximation we made 
is not good at  $\mu\sim m_{e}$. However, the 
contribution of this region is small, and the corresponding 
errors  are not big.

\section{Calculation of $dP/dxdL$}
The essential ingredient of Eq. (\ref{eq:20}) is the probability distribution
$dP/dx dL$ for the photon emission in the electromagnetic field of the
electrosphere. 
Due to presence of the product $n_{F}(E)[1-n_{F}(E')]$ in (\ref{eq:20})
the emissitivity is dominated by the photon emission from the electrons
near the Fermi surface with $p\sim \mu\gg m_{e}$. This allows one to use
for the photon spectrum $dP/dx dL$ the quasiclassical relativistic formulas.
In this work we evaluate this  spectrum  
within the LCPI formalism \cite{Z1,Z_YAF98}. In this approach it can be
written as
\beq
\frac{d P}{d
xdL}=2\mbox{Re}
\int\limits_{0}^{\infty}d
\xi
\hat{g}(x)\left[{\cal
K}(\r_{2},\xi|\r_{1},0)
-{\cal
K}_{v}(\r_{2},\xi|\r_{1},0)
\right]
\Big{|}_{\r_{1}=\r_{2}=0}.
\label{eq:30}
\eeq
Here $\hat g$ is the spin vertex operator given by
\beq
\hat{g}(x)=\frac{g_{1}(x)}{M^{2}(x)}\frac{\partial }{\partial \r_{1}}\cdot
\frac{\partial }{\partial \r_{2}}+g_{2}(x)
\label{eq:31}
\eeq
with 
$g_{1}(x)=\alpha(1-x+x^{2}/2)/x$ and
$g_{2}(x)=\alpha m_{e}^{2}x^{3}/2M^{2}(x)$,
$M(x)=px(1-x)$,
${\cal K}$ is the Green's function for a 
two-dimensional Schr\"odinger equation with the Hamiltonian
\beq
\hat{H}=-\frac{1}{2M(x)}
\left(\frac{\partial}{\partial \r}\right)^{2}
+v(\r)+\frac{1}{L_{0}}\,,
\label{eq:40}
\eeq
where 
$L_{0}=2M(x)/\epsilon^{2}$,
$\epsilon^{2}=m_{e}^{2}x^{2}+(1-x)m_{\gamma}^{2}$ 
($m_{\gamma}$ is the photon quasiparticle mass),
the form of the potential $v$ 
will be given below.
In (\ref{eq:30})-(\ref{eq:40}) $\r$ is the coordinate transverse to the 
electron momentum $\pb$, the longitudinal (along $\pb$) coordinate $\xi$
plays the role of time.
The ${\cal K}_{v}$ in (\ref{eq:30}) is the free Green's function for $v=0$.
Note that at low density and vanishing mean field the quantity $L_{0}$ 
coincides with the real photon formation length $l_{f}$ \cite{Z1}
which characterizes the dominating scale in the $\xi$-integration on the 
right-hand side of (\ref{eq:30}).

The potential in the Hamiltonian (\ref{eq:40}) 
can be written as $v=v_{m}+v_{f}$. The terms  
$v_{m}$ and $v_{f}$ correspond to the mean and fluctuating
components of the vector potential of the electron gas.
Note that when $l_{f}$ is small compared to the scale of variation 
of $\mu$ (along the electron momentum)
one can neglect the $\xi$-dependence of the potential $v$
in evaluating $dP/dxdL$. 
The mean field component is purely real
$v_{m}=-x\fb\!\cdot\!\r$ with $\fb=e \partial V/\partial \r$ 
\cite{Z_YAF98,Z_synch}. 
It is related to the transverse force from the mean field.
Note that, 
similarly to the classical radiation 
\cite{LL2}, the effect of the longitudinal force 
along the electron momentum $\pb$ is suppressed by a factor 
$\sim (m_{e}/E)^{2}$, and can be safely neglected.
The term $v_{f}$ 
can be evaluated similarly to the case 
of the quark-gluon plasma discussed in \cite{AZ}.  
This part
is purely imaginary 
$v_{f}(\r)=-i P(x\r)$, where
\beq
P(\r)=e^{2}\int\limits_{-\infty}^{\infty} d\xi 
[G(\xi,0_{\perp},\xi)-G(\xi,\r,\xi)]\,,
\label{eq:50}
\eeq
$
G(x-y)= 
u_{\mu}u_{\nu} D^{\mu\nu}
$,
$
D^{\mu\nu}={\Large\langle}
A^{\mu}(x)A^{\nu}(y)
{\Large\rangle}
$
is the correlation function of the electromagnetic potential
(the mean field is assumed to be subtracted)
in the electron plasma, $u_{\mu}=(1,0,0,-1)$ is the light-cone 4-vector
along the electron momentum. 
Note that the function $P(\r)$ is gauge invariant
by construction, and one can use $D^{\mu\nu}$ in any gauge.
The formula (\ref{eq:50}) may be rewritten
as (below we replace the argument of $P(\r)$ 
by $\rho=|\r|$ since $P(\r)$ does not depend on the direction of 
the vector $\r$)
\beq
P(\rho)=\frac{e^{2}}{(2\pi)^{2}}\int
d\qbt [1-\exp(i\qbt\r)]D(\qbt)\,,
\label{eq:60}
\eeq
where the function $D$ in terms of the correlator $G$ in momentum
representation reads 
\beq
D(\qbt)=\frac{1}{2\pi}\int\limits_{-\infty}^{\infty}
dq_{0} dq_{z}\delta(q_{0}-q_{z})
G(q_{0},\qb_{\perp},q_{z})\,.
\label{eq:61}
\eeq

The function $D(\qbt)$ may be expressed in terms
of the longitudinal and transverse photon self energies 
$\Pi_{L,T}$.
We use for them the formulas of the hard dense loop approximation (HDL)
\cite{HDL1,HDL2}.  
The details of the calculations are given in the Appendix A.

The function $P(\rho)$ has been introduced, for the first time,
in the context of the problem
of propagation of relativistic positroniums through 
amorphous media \cite{Z_A2e}, where the atomic size plays the role of 
the inverse Debye mass.
In our approach the function $P(\rho)$ contains all information about
the electron-electron interaction which is necessary for description
of multiple scattering of a given electron in the fluctuating electromagnetic 
field generated by other electrons. In particular,
all the Pauli-blocking effects in the process of electron 
multiple scattering are automatically accumulated in $P(\rho)$.
It is worth noting that in the approximation of static Debye screened 
scattering centers the function $P(\rho)$ reduces to 
$n \sigma(\rho)/2$ \cite{AZ}, 
where $n$ is the number density of the medium, and 
\beq
\sigma(\rho)=8\alpha^{2}\int d\qb
\frac{[1-\exp(i\qb\r)]}{(\qb^{2}+m_{D}^{2})^{2}}
=\frac{8\pi \alpha^{2}}{m_{D}^{2}}
\left[1-\rho m_{D}K_{1}(\rho m_{D})\right]
\label{eq:91}
\eeq
is the well known dipole cross section for scattering of an
$e^{+}e^{-}$ pair of size $\rho$ on the Debye screened  scattering center
(in (\ref{eq:91}) $K_{1}$ is the Bessel function). 
In the static approximation at $\rho\ll 1/m_{D}$ 
one can obtain from (\ref{eq:91})  $P(\rho)\approx n C\rho^{2}/2$,
where $C\approx 4\pi\alpha^{2}\ln(2/\rho m_{D})$ is a smooth function
of $\rho$. In the limit $\rho\ll 1/m_{D}$
the function $P(\rho)$ in the HDL approximation
also becomes almost quadratic.

The quadratic approximation $P(\rho)\propto \rho^{2}$ 
in the LCPI approach 
is equivalent to the Fokker-Planck approximation in Migdal's approach
\cite{Z_YAF98}.
It is not very accurate but reasonable for bremsstrahlung in 
ordinary materials. In this case
the dominating $\rho$-scale is $\sim 1/m_{e}x$, and the spectrum 
is controlled by behavior of $P(\rho)$ at the scale $\sim 1/m_{e}$
which is much smaller than the screening radius $\sim 1/\alpha m_{e}Z^{1/3}$
(here $Z$ is the atomic number).
In the case of the relativistic electron gas the situation
is quite different. Now, in the dominating $\rho$-region,
the argument of $P(\rho)$ is $\sim (0.1\div 2)/m_{D}$.
In this region $P(\rho)$ is essentially  non-quadratic.
It is well seen from Fig.~1a, in which we plot the results of numerical 
calculations of $P(\rho)$ for several values 
of the ratio $T/m_{D}$. The results are presented in a dimensionless
form. For comparison in Fig.~1a we also show the predictions of the static
approximation at $T=0$ (when
$m_{D}=\mu\sqrt{4\alpha/\pi}$) 
obtained with the dipole cross section (\ref{eq:91}). 
One can see that at $\rho\sim (0.1\div  2)/m_{D}$ the function $P(\rho)$ 
is almost linear in $\rho$.

To demonstrate the relative effect
of the longitudinal and transverse modes in Figs.~1b,c we show separately
the contributions related to $\Pi_{L}$ and $\Pi_{T}$. 
One sees that at $\rho\lsim 1/m_{D}$ the L and T contributions are close
to each other. However, at $\rho\gsim 2/m_{D}$ the longitudinal
part flattens, while the transverse magnetic one continues to grow
(for $T/m_{D}$ not very close to zero).
This growth of the transverse part is a consequence of the well known 
absence of the static magnetic screening in the electron plasma.
Note, however, that from the point of view of the photon emission
the growth of the magnetic contribution with $\rho$ is not important
since the photon spectrum is dominated 
by $\rho\lsim 1/\epsilon \sim 1/m_{D}$.

The growth of $P(\rho)$ with temperature is due to the presence of 
the Bose-Einstein factor in the function $D$ (\ref{eq:70}). 
From Fig.~1a one can see that
the prediction of the HDL approximation
at $T\ll m_{D}$, similarly to the static model, 
flatten at $\rho\gsim 2/m_{D}$. 
However, the static 
model exceeds the HDL prediction
by a factor $\sim 2.5$.
The fact that at $T=0$ the static approximation overestimates
$P(\rho)$ is quite natural, since the Pauli-blocking effects 
reduce the effective number of the scatterers. Note, however, that 
it would be incorrect to interpret the growth of 
$P(\rho)$ with temperature as an artifact associated only with 
the decrease of the Pauli-blocking at high temperatures. 
The function $P(\rho)$ in the HDL approximation accumulate all the collective
effects in the soft modes of electromagnetic field in the electron 
plasma at the momentum scale $\sim m_{D}\ll \mu$. 
In particular, it accounts 
for the temperature dependence of the density of the plasmon 
excitations. Note that, physically, the appearance of 
$P(\rho)$ is due to Landau
damping of the longitudinal and transverse modes.   

It is worth noting that the collective effects cannot be accounted for
consistently in the naive
modification of the photon propagator in the amplitude of elastic 
$e^{-}e^{-}\to e^{-}e^{-}$
scattering as was assumed in \cite{Gale}. One of the consequence of
the inadequacy of this prescription is a strong overestimate of the magnetic
contribution in \cite{Gale}. It is connected with the 
$1/\theta^{4}$ ($\theta$ is 
the scattering angle) behavior of the magnetic contribution to elastic
$e^{-}e^{-}\to e^{-}e^{-}$ cross section. To perform the $\theta$-integration 
the authors of \cite{Gale} introduced some minimal momentum transfer.
In contrast to \cite{Gale} the magnetic contribution to the function
$D(\qb_{\perp})$ is $\propto 1/\qb^{2}_{\perp}$ at 
$\qb_{\perp}\to 0$\footnote{The same
occurs in the hard thermal loop
approximation for a hot relativistic plasma with zero chemical
potential \cite{AGZ}. Note, however, that a very elegant formula
for the analog of our function $D(\qb_{\perp})$ obtained in \cite{AGZ}
is not valid for a strongly degenerate electron plasma.}, and the
$\qb_{\perp}$-integration
in the formula for $P(\rho)$ (\ref{eq:60}) converges at small $\qb_{\perp}$. 
This change in
the small angle behavior of the magnetic contribution in the prescription
of \cite{Gale} and in our approach is connected with the dynamical
magnetic screening which was not consistently 
accounted for in \cite{Gale}.
In principle, physically, it is evident 
that the concept of the elastic $e^{-}e^{-}\to e^{-}e^{-}$ amplitude
itself is ill-defined for the momentum transfer $\lsim m_{D}$, 
where the collective effects come into play.

Note that in terms of $P(\rho)$
the transverse momentum broadening distribution
of an electron propagating a distance $L$ through the electron gas  can 
be written as \cite{Z_A2e}
\beq
I(\qbt)=\frac{1}{(2\pi)^{2}}
\int d\r \exp{[i\qbt\r - L P(\rho)]}\,.
\label{eq:92}
\eeq
This formula looks like the prediction of the eikonal approximation which
neglects the variation of the electron tranverse coordinate. However,
the path integral calculations \cite{Z_A2e} show that it is valid beyond
the eikonal approximation as well.

Let us  turn to calculation of the spectrum with the help of (\ref{eq:30}).
Treating $v_{f}$ as a perturbation one can write
\beq
{\cal K}(\xi_2,\r_{2}|\xi_1,\r_{1})=
{\cal K}_{m}(\xi_2,\r_{2}|\xi_1,\r_{1})
-i\int d\xi d\r
{\cal K}_{m}
(\xi_2,\r_{2}|\xi,\r)
v_{f}(\r){\cal K}_{m}(\xi,\r|\xi_1,\r_{1})+\dots\,\,,
\label{eq:100}
\eeq
where ${\cal K}_{m}$ is the Green's function for $v_{f}=0$.
Then (\ref{eq:30}) can be written as
\beq
\frac{d P}{d
xdL}=\frac{d P_{m}}{d
xdL}+
\frac{d P_{f}}{d
xdL}\,.
\label{eq:110}
\eeq
Here the first term on the right-hand side comes from the
${\cal K}_{m}-{\cal K}_{v}$ in (\ref{eq:30}) after representing
${\cal K}$
in the form (\ref{eq:100}). It corresponds to the photon emission
in a smooth mean field.  The second term 
comes from the series in $v_{f}$ in (\ref{eq:100}). This term can be viewed
as the radiation rate due to electron multiple scattering in the fluctuating
field in the presence of a smooth external field.

The analytical expression for the Green's function 
for the Hamiltonian with a constant force
is known (see, for example \cite{FH}). In our case the
${\cal K}_{m}$ can be written as
\beq
{\cal K}_{m}(\xi_2,\r_{2}|\xi_1,\r_{1})
\!=\!\frac{M}{2\pi i\xi}
\exp
\left\{i\left[
\frac{M(\r_{2}-\r_{1})^{2}}{2 \xi}
-\frac{x \xi \fb\cdot(\r_{2}+\r_{1})}{2}
-\frac{x^{2}\fb^{2} \xi^{3}}{24 M}
-\frac{\xi}{L_{0}}
\right]
\right\}
\label{eq:111}
\eeq
with $\xi=\xi_{2}-\xi_{1}$.
With this expression from (\ref{eq:30}) after simple calculations
one can obtain a spectrum similar to the well known quasiclassical 
synchrotron spectrum \cite{BK}
which can be written in terms of the Airy function
$\mbox{Ai}(z)=\frac{1}{\pi}\sqrt{\frac{z}{3}}K_{1/3}(2z^{3/2}/3)$ (here
$K_{1/3}$ is the Bessel function). 
In the case of interest, for a nonzero photon
quasiparticle mass it reads \cite{Z_synch}
\beq
\frac{dP_{m}}{dxdL}=
\frac{a}{\kappa}\mbox{Ai}^{'}(\kappa)+
b\int_{\kappa}^{\infty}dy\mbox{Ai}(y)\,,
\label{eq:120}
\eeq
where 
$a=-{2\epsilon^{2}g_{1}}/{M}$,
$
b=M  g_{2}-{\epsilon^{2}g_{1}}/{M}
$,
$\kappa=\epsilon^{2}/(M^{2}x^{2}\fb^{2})^{1/3}$.
Inspecting the longitudinal integrals for the photon radiation
in an external field one can find that 
the effective photon formation length for the 
mean field mechanism is given by 
$\bar{L}_{m}\sim \mbox{min}(L_{0},L_{m})$, where 
$L_{m}=(24M/x^{2}\fb^{2})^{1/3}$ \cite{Z_synch}. 
A similar estimate can be obtained from the criterion of separation
of the photon and electron wave packets.
Note that the analytical expression for the Green's function 
for the oscillator with a constant force 
is known as well (see \cite{FH}). Making use of this Green's function
one can obtain for $P(\rho)\propto \rho^{2}$
the radiation rate in the form given 
in \cite{BT}, where
Migdal's approach within the Fokker-Planck approximation
was generalized to the case with an external field.
The formulas of \cite{BT} were used in \cite{Harko2}.
However, as already noted the approximation $P(\rho)\propto \rho^{2}$
is clearly not adequate for the electrosphere.

Let us discuss now the fluctuation component ${dP_{f}}/{dxdL}$. 
We represent it in the form
\beq
\frac{d P_{f}}{d
xdL}=
\frac{d P_{f}^{BH}}{d x}+\frac{d P_{f}^{LPM}}{d
x}\,,
\label{eq:121}
\eeq
where the first term on the right-hand side corresponds to the leading 
order in expansion in $v_{f}$ in (\ref{eq:100}), and the second one
to the sum of the higher order terms. The $dP_{f}^{BH}/dxdL$ is the analog of 
the Bethe-Heitler spectrum in ordinary materials, while the 
$dP_{f}^{LPM}/dxdL$ describes the LPM correction.  
For the Bethe-Heitler term one can obtain from (\ref{eq:30}), (\ref{eq:100})
\beq
\frac{d P_{f}^{BH}}{d
x}=
2\int
d\r\,
 W(x,\r,\fb)
P(\rho x)\,,
\label{eq:122}
\eeq
\beq
W(x,\r,\fb)=-\mbox{Re}\,\,\hat{g}(x)
\Phi(x,\r,\r_{1},\fb)
\Phi(x,\r,\r_{2},\fb)\Big|_{\r_{1}=\r_{2}=0}\,,
\label{eq:123}
\eeq
\beq
\Phi(x,\r,\r',\fb)
=
\int_{-\infty}^{0} d\xi {\cal K}_{m}(\r,0|\r',\xi)\,.
\label{eq:124}
\eeq
Note that for a nonzero $\fb$ the function $W$
cannot be viewed as a probability density 
for the $|\gamma e\rangle$ Fock component of the physical photon
(it is even not positively defined).
This is connected with the fact that in an external field
the $|\gamma e\rangle $ Fock component is not stable, and decays through the tunnel
transition into free photon and electron.
The analog of the representation for the LPM correction 
derived in \cite{Z_SLAC1} for nonzero 
mean field reads
\bea
\frac{d P_{f}^{LPM}}{d
x}=2\mbox{Re}\,\hat{g}(x)
\int\limits_{0}^{\infty}
d\xi
\int
d\r\,
\Phi(x,\r,\r_{2},\fb)
P(\rho
x)
\tilde{\Phi}(x,\r,\r_{1},\fb,\xi)
\Big|_{\r_{1}=\r_{2}=0}
\,,
\label{eq:125}
\eea
where the
function
$\tilde{\Phi}(x,\r,\r_{1},\fb,\xi)$ is
the solution of
the two-dimensional Schr\"odinger equation with the Hamiltonian (\ref{eq:40}).
The
boundary
condition for
$\tilde{\Phi}(x,\r,\r_{1},\fb,\xi)$
is
$
\tilde{\Phi}(x,\r,\r_{1},\fb,0)=
\Phi(x,\r,\r_{1},\fb)P(\rho
x)\,.
$

In the case of zero $\fb$ the function $W$ may be written
as a density for the  $|\gamma e\rangle $ Fock state
\beq
W(x,\r)=
\frac{1}{2}
\sum\limits_{\{\lambda_{i}\}}
|\Psi(x,\r,\{\lambda_{i}\})|^{2}\,,
\label{eq:126}
\eeq
where $\Psi(x,\r,\{\lambda_{i}\})$
is the light-cone wave function for the $e\to \gamma e'$
transition, $\{\lambda_{i}\}=(\lambda_{e},\lambda_{e'},\lambda_{\gamma})$ 
a set of helicities. Note that contrary to the case $\fb\neq 0$ now,
due to the azimuthal symmetry of the Hamiltonian, 
the light-cone wave functions have
definite azimuthal quantum numbers.
The LPM correction in this case also can be written in terms of 
the light-cone wave functions. The results is similar to that
for ordinary materials \cite{Z_SLAC1,Z_YAF98} 
\bea
\frac{d P_{f}^{LPM}}{d
x}=-\mbox{Re}
\sum\limits_{\{\lambda_{i}\}}
\int\limits_{0}^{\infty}
d\xi
\int
d\r\,
\Psi^{*}(x,\r,\{\lambda_{i}\})
P(\rho
x)
\tilde{\Phi}(x,\r,\{\lambda_{i}\},\xi)
\,.
\label{eq:127}
\eea
The boundary condition 
for
$\tilde{\Phi}(x,\r,\{\lambda_{i}\},\xi)$ is now
$
\tilde{\Phi}(x,\r,\{\lambda_{i}\},0)=
\Psi(x,\r,\{\lambda_{i}\})P(\rho
x)\,.
$
The light-cone wave functions appear in the formulas 
(\ref{eq:126}), (\ref{eq:127}) 
from the $\xi$-integrals in
(\ref{eq:30}) and (\ref{eq:100}) of the Green's function 
${\cal K}_{m}$ and action
of the vertex operator written in terms of the helicity projectors
as was done in \cite{AZ}.

The formulas for the light-cone wave functions are given
in the Appendix B. 
Making use of the formulas given there
one can obtain for the probability distribution $W$ for the 
$e\to \gamma e'$
transition at $\fb=0$
\beq
 W(x,\r)=
\frac{\alpha}{2\pi^{2}}\left\{
\frac{[1+(1-x)^{2}]}{x}\epsilon^{2}K_{1}^{2}(\rho\epsilon)
+x^{3}m_{e}^{2}K_{0}^{2}(\rho\epsilon)\right\}\,,
\label{eq:140}
\eeq
where $K_{0,1}$ are the Bessel functions.
Due to exponential decrease of  $K_{0,1}$ in (\ref{eq:140})
the dominating $\rho$ scale in the formula (\ref{eq:122}) for the 
fluctuation term is $\sim 1/\epsilon$.

For nonzero $\fb$ the azimuthal symmetry is absent.
This makes the problem considerably more complicated.
In the present work we have first calculated the spectrum ${dP_{f}}/{dxdL}$
for $\fb=0$.
We observed that the LPM correction in (\ref{eq:121}) is negligible
as compared to the Bethe-Heitler term.
Also, the Bethe-Heitler term itself turns out to be much smaller than the
mean field term $dP_{m}/dxdL$. It is clear that a nonzero $\fb$ 
will make $dP_{f}/dxdL$ even smaller.
For this reason an accurate calculation of the fluctuation term for nonzero
$\fb$ does not make much sense. 
We have taken into account
the effect of the transverse force using qualitative arguments based 
on the estimates of the coherence lengths with and without tranverse force.
The mean field should suppress the coherence length.
The suppression of the radiation rate should be approximately
the same \cite{GG}. Thus, the mean field suppression factor can be written
as the ratio of the formation lengths with and without the mean field.
The coherence length in the presence of the mean field is $\sim \bar{L}_{m}$.
Without the mean field in the regime of weak LPM suppression
the coherence length is given by $L_{0}$. So one has the mean field
suppression factor
$S_{m}\approx\bar{L}_{m}/L_{0}$.
Note that due to reduction of the effective formation length 
the LPM effect  should become even smaller for
a nonzero mean field. 

To illustrate the relative contributions of the mean field and fluctuation
mechanisms to $dP/dxdL$ we plot them in Fig.~2 
for $\mu=10$ MeV and $T=0.2$ and $T=1$ MeV. 
The mean field part shown in Fig.~2 corresponds to the spectrum 
averaged over all directions of the electron momentum.
The fluctuation contribution
has been calculated without the mean field suppression factor.
Note that we perform calculations with 
the $k$-dependent
photon quasiparticle mass extracted from the relation
$m^{2}_{\gamma}=\Pi_{T}(\sqrt{k^{2}+m^{2}_{\gamma}},k)$
\footnote{
We ignore the influence of the medium
effects on $m_{e}$ \cite{me} since the photon bremsstrahlung
in the region $x\ll 1$, which dominates the emissitivity,  
is not very sensitive to the electron quasiparticle mass.}.
This gives
$m_{\gamma}$ rising from $m_{D}/\sqrt{3}$ at $k\ll m_{D}$ to $m_{D}/\sqrt{2}$
at $k\gg m_{D}$
with the Debye mass 
$m_{D}^{2}=\frac{4\alpha}{\pi}(\mu^{2}+\pi^{2}T^{2}/3)$.
From Fig.~2 one sees that the fluctuation contribution is suppressed by 
a factor $\sim 10^{-2}$. To illustrate the role of the 
finite photon quasiparticle mass we presented in Fig.~2 the results for 
zero $m_{\gamma}$ as well (thin curves). It is seen that the photon mass
suppression (called usually the Ter-Mikaelian effect) is very strong at
small $x$. The effect is especially dramatic for the fluctuation part
where the well known $1/x$ form of the spectrum is changed into $\propto x$. 
This effect was ignored in the analyses \cite{Zh1,Harko2} where
the massless formulas  have been used. The results shown in Fig.~2 indicate
clearly that the massless approximation 
is inadequate. 

As previously mentioned, our calculations show that for the fluctuation
mechanism the LPM suppression is negligible. This is in a
contradiction with the  analysis \cite{Gale} where the authors find
a very strong LPM suppression (about  $\sim 1/300$ at the 
photon momentum $k=0.5$ MeV
for electron energy $10$ MeV). For calculation of the 
LPM suppression the authors of
\cite{Gale} have used Migdal's 
formulas with zero photon mass putting there $Z=1$ .
However, one can easily show that Migdal's formulas become inadequate
for the electrosphere. We explain this in the language of the LCPI approach. 
Migdal's approach \cite{M} corresponds in the LCPI formalism to 
quadratic parametrization $P(\rho)\approx nC\rho^{2}/2$.
As described above, this approximation is not accurate for electrosphere,
but nevertheless it is suitable for our qualitative analysis.
In the quadratic approximation the Hamiltonian (\ref{eq:40}) 
takes the oscillator
form with $\Omega=\sqrt{-in C x^{2}/M(x)}$. 
The LPM suppression
factor, $S_{LPM}$, can be written in terms of the dimensionless parameter 
$\eta=|\Omega|L_{0}$ \cite{Z1,Z_YAF98}. 
The LPM suppression becomes strong
at $\eta\gg 1$. In this limit $S_{LPM}\approx 3/\eta\sqrt{2}$ \cite{Z1}.
The LPM effect is negligible for 
$\eta\ll 1$ when $S_{LPM}(\eta)\approx 1-16\eta^{4}/21$ \cite{Z1},
Note that even at $\eta\sim 1$ the LPM suppression is relatively small since 
$S_{LPM}(1)\approx 0.86$. 
A very strong suppression obtained in \cite{Gale} is mostly due to 
the neglect of the photon mass. The finite photon mass 
reduces strongly the 
$L_{0}$ and correspondingly the parameter $\eta$ (about a factor $\sim 400$
for $k= 0.5$ and $p\sim 10$ MeV).  Also, for the electrosphere
there is no the well known large Coulomb logarithm 
$\ln(1/\alpha)\sim 5$ (which comes from the logarithm in 
the dipole cross section \cite{Z_SLAC2})
in the $|\Omega|$,
which is present in Migdal's
formulas derived for ordinary materials.
Both these effects reduce drastically the value of $\eta$ for
the electrosphere as compared to that in Migdal's approach.
As a result, the LPM suppression in the electrosphere 
turns out to be negligible.

\section{Numerical results and discussion}
In this section we present the numerical results for the emissitivity
and radiated energy flux.
The results were obtained with some modification of the spectrum
$dP/dxdL$ in the noncollinear region. 
As we mentioned earlier, the collinear approximation we use   
becomes invalid
for very soft photons with $k\lsim m_{\gamma}$. 
In this region the formalisms \cite{AMY1,AZ,Z1} do not apply.
In particular, the LCPI approach \cite{Z1},
which assumes that the transverse momentum integration
comes up to infinity, should overestimate the photon spectrum at
$k\lsim m_{\gamma}$. 
To take into account this effect (at least, qualitatively) 
in calculating the radiated 
energy flux we multiplied $dP/dxdL$ by the kinematical suppression factor 
$S_{kin}(k)=1-\exp(-k^{2}/m_{\gamma}^{2})$. This factor
does not give a big effect. It
suppresses the radiated energy by $\sim 10-15$\%
at $T\sim 0.1\div 0.2$ MeV and  
$\sim 1-2$\%
at $T\sim 1$ MeV. This says that the errors from
the noncollinear configurations
are small.

In Fig.~3 we show the emissitivity for $\mu=5$ and $\mu=10$ MeV
evaluated for $T=0.2$ and $T=1$ MeV as a function of $\omega$.
One sees that the contribution of the mean field emission (thick solid line)
exceeds the fluctuation emission without mean field suppression
(dashes) by a factor $\sim 10^{2}$. The mean field suppression
gives additional reduction of the fluctuation contribution (thin solid line)
by a factor $\sim 3-4$. Note that in our quasiclassical  approximation
at a given $\mu$ there is no photon emission at $\omega<\omega_{p}^{e}$.
For this reason the differential emissitivity shown in Fig.~3 
vanishes abruptly at $\omega=\omega_{p}^{e}=m_{\gamma}(k=0)$.
From Fig.~3 one can see that, despite the Pauli-blocking suppression,
even at $T=0.2$ MeV the contribution of energetic photons with energy
about several units of $\omega_{p}^{e}$ is important. This demonstrates
that the restriction on the photon energy 
$\omega<\sqrt{\omega_{p}^{e\,2}+m_{e}^{2}}$ imposed by the authors
of \cite{Gale} is clearly inadequate.

In Fig.~4 we plot the differential
radiated energy flux $dF/d\omega$ for
$\mu(0)=10$ and $\mu(0)=20$ MeV obtained for
$T=0.2$ and $T=1$ MeV.
For the fluctuation contribution we show the results with and without
the mean field suppression factor $S_{m}$. 
For comparison the black body spectrum is also shown. 
The mean Coulomb field of the electrosphere reduces the fluctuation term 
by a factor $\sim 3-4$.
From Fig.~3,~4 one can see that the relative contribution
of the fluctuation mechanism is very small compared to the
mean field emission. Thus, in some sense
we have a situation similar to that for photon radiation from an atom 
with large $Z$.
Note that the form of the spectrum for the 
mean field mechanism is qualitatively similar to that 
for the black body radiation.

In Fig.~5 we show the total energy flux 
$
F=\int_{0}^{\infty}d\omega {dF}/{d\omega}
$
scaled to the black body radiation
as a function of temperature. 
For comparison we also plot the predictions 
for bremsstrahlung obtained in \cite{Gale,Zh1,Harko2}.
We also show there the energy flux from
the $e^{+}e^{-}$ pair production \cite{Usov1,Usov2}. We define it 
as
\beq
F_{\pm}=\int_{0}^{h_{max}}dh Q_{\pm}(h)\approx
\sqrt{\frac{3\pi}{2\alpha}}\int_{\mu_{min}}^{\mu(0)}\frac{d\mu}{\mu^{2}}
Q_{\pm}(h(\mu))\,.
\label{eq:170}
\eeq
Here $Q_{\pm}$ is the energy flux from $e^{+}e^{-}$ pairs 
per unit time and volume. We write it as in 
\cite{Usov1,Usov2} 
$
Q_{\pm}=E_{e^{+}e^{-}}dN_{e^{+}e^{-}}/dtdV
$, where $E_{e^{+}e^{-}}\approx 2(m_{e}+T)$ is the typical energy of 
$e^{+}e^{-}$ pairs, and $dN_{e^{+}e^{-}}/dtdV$ the rate of 
$e^{+}e^{-}$ pair production per unit time and volume given by
\beq
\frac{dN_{e^{+}e^{-}}}{dtdV}\approx \frac{3 T^{3}\mu}{2\pi^{3}}
\sqrt{\frac{\alpha}{\pi}}\exp{\left(-\frac{2m_{e}}{T}\right)}J(\xi)
\label{eq:171}
\eeq
with $\xi=\frac{2\mu}{T}\sqrt{\frac{\alpha}{\pi}}$, and the function 
$J$ is defined as in
\cite{Usov2}
$$
J(x)=\frac{x^{3}\ln{(1+2/x)}}{3(1+0.074x)^{3}}
+\frac{\pi^{5}x^{4}}{6(13.9+x)^{4}}\,.
$$
From Fig.~5 one sees that in the region $T\sim 0.1\div 1$ MeV the mean 
field photon emission exceeds considerably both the 
fluctuation bremsstrahlung and the energy flux from $e^{+}e^{-}$ pair
production.

Figs.~4,~5  demonstrate that the energy flux from the mean 
field photon emission
may be of the same order of magnitude as the black body radiation.
It says that the approximation of optically thin electrosphere is not
very good, and the photon absorption and stimulated emission may be important.
However, since the radiation rate we obtained does not exceed 
the black body limit,  
they  can not modify strongly our
results. Note that the authors of \cite{Harko2} obtained for 
$T\lsim 1$ MeV the energy flux
considerably exceeding the black body limit. This can be seen
from Fig.~5, where the results of \cite{Harko2} 
at $\mu(0)=20$ MeV are shown.
The authors of \cite{Harko2} claim
that the electrosphere may radiate stronger than a black body. 
This statement is obviously 
incorrect. The violation of the black body limit in 
\cite{Harko2} is just a signal that the thin medium approximation 
becomes inadequate at high emissitivity. As far as a very large emissitivity
obtained in \cite{Harko2} is concerned, as we already mentioned, 
it may be due to incorrect description
of the Pauli-blocking and neglect of the photon mass.

As we mentioned earlier,
our assumption that the photon emission is a local process
is valid if $l_{f}\sim \bar{L}_{m}\ll L_{el}$, 
where $ L_{el}$ is the typical scale of
variation of the potential $v_{m}$ along the electron trajectory. 
For the chemical potential (\ref{eq:10})
it can be defined as $L_{el}\sim H\mu(0)/\mu(h) \cos{\theta}$,
where $\theta$ is the angle between the electron momentum and the star surface
normal. Evidently, the contribution of the configurations
with $\bar{L}_{m}\gsim L_{el}$ into the photon spectrum 
will be suppressed by the finite-size suppression factor $S_{fs}\sim
\mbox{min}(L_{el},\bar{L}_{m})/\bar{L}_{m}$. We have checked numerically 
that this suppression factor gives a negligible effect. This justifies
the local approximation.

According to the simulation of the thermal evolution of young
quark stars performed in \cite{Usov-LC} the temperature at the 
star's surface becomes $\sim 0.2$ MeV at $t\sim 1$ s. 
However, in the analysis \cite{Usov-LC} the mean field bremsstrahlung
was not taken into account. 
In the light of our results
one can expect that the cooling of the bare quark star's surface  
should go somewhat faster than predicted in 
\cite{Usov-LC}. It is worth noting
that in the initial stage of the quark star evolution 
the mean field photon emission can only modify the temperature
near the star surface. 
The evolution of the star core 
temperature is driven by the neutrino emission \cite{Usov-LC}
since for an extended period of time the neutrino luminosity is much 
larger than the photon (and
$e^{+}e^{-}$) luminosity \cite{Usov-LC}.
Higher luminosity 
due to the mean field bremsstrahlung
increases the possibility for detecting bare quark stars.   
From the point of view of the light curves at $t\gsim 1$ s it would 
be interesting to investigate the mean field bremsstrahlung
for $T\lsim 0.1$ MeV as well. 
However, at such temperatures the photon emission from the
nonrelativistic region of the electrosphere may be important,
where our formulas become inapplicable. As far as the contribution of the
relativistic region $\mu\gg m_{e}$ is concerned. Extrapolation of the
curves shown in Fig.~5 to 
$T\lsim 0.1$ MeV allows one to expect that 
the mean field emission will dominate the energy flux at lower temperatures
as well.

A remark is in order here on the photon distribution seen by a distant 
observer.
For obtained values of the energy flux the radiation cannot stream outward 
freely. The point is that near the star surface 
the thermalization time in the comoving frame for the 
$e^{+}e^{-}\gamma$ wind is negligibly small as compared
to the star radius.  This follows from estimates of the mean free
path, $\lambda$, related to $\gamma+e^{\pm}\rightarrow \gamma+e^{\pm}$ and 
$\gamma+\gamma\leftrightarrow e^{+}+e^{-}$ processes. The qualitative 
calculations give 
$\lambda\sim 10^{-3}$cm at $T\sim 0.1$ MeV and $\lambda\sim 10^{-6}$cm
at $T\sim 1$ MeV. For this reason the $e^{+}e^{-}\gamma$ wind can be described
as a hydrodynamical flow. The hydrodynamical description is valid
up the the freezeout surface, 
beyond which the radiation streams outward almost
freely. For an observer at large distance from the star the photon spectrum
is close to the black body one with a temperature
$T_{ext}= T_{fr}\Gamma_{fr}$, where $T_{fr}$ is the wind temperature 
and $\Gamma_{fr}$ the bulk Lorentz factor of the wind
at the freezeout level \cite{Pacz,Gr}. One can show that for a 
relativistic wind 
$T_{fr} \Gamma_{fr}\approx T_{i} \Gamma_{i}$
 \cite{Pacz,Gr}, where $T_{i}$ is the wind temperature after 
its thermalization and $\Gamma_{i}$ the bulk Lorentz factor of the wind
near the star surface. 
For $T\sim 0.1$ MeV the electron fraction in the $e^{+}e^{-}\gamma$ wind
after thermalization is small.
Simple qualitative calculations give in this case  
$T_{i}\Gamma_{i}\approx T (3\kappa\Gamma_{i}^{2}/16)^{1/4}$, 
where $\kappa=(F+F_{\pm})/F_{bb}$.
As a plausible estimates one can take
$\Gamma_{i}^{2}\sim 3$ and $\kappa\sim 1$. Then one obtains $T_{ext}\sim
0.85 T$.
For $T\sim 1$ MeV the electron fraction in the wind after 
thermalization becomes close to that for relativistic plasma.
In this case $T_{i}\Gamma_{i}\approx T (3\kappa\Gamma_{i}^{2}/44)^{1/4}$.
Taking  $\kappa\sim 0.4$ one obtains
$T_{ext}\sim 0.5 T$.
Note that in both the cases beyond the freezeout surface  
the fraction of $e^{\pm}$ pairs in the wind is negligibly small \cite{Gr}.

Note that our calculations probably do not apply to quark stars
in the color flavor locked (CFL) superconducting phase.
Previously it was suggested \cite{Usov-CFL} that, despite the absence
of electrons in the bulk SQM in the CFL phase,
the electrosphere may exist due to the surface quark charge
\cite{Madsen}.
However, the recent analysis \cite{CFL-el} gives evidence in favor of absence
of such a surface charge.
But for the CFL phase may exist a significant photon emission from the SQM
itself due to the photon-gluon mixing \cite{Rapp}.
The results of \cite{Rapp} show that this radiation is 
comparable to the black body limit.
Since we also obtain the radiation rate comparable to the black body 
radiation it may be difficult to distinguish a
bare quark star in the CFL phase from that in normal 
(or 2SC) phase.

\section{Conclusion}
In summary, we have evaluated the photon emission from the electrosphere
of a bare quark star (in normal or 2SC phase).  
The analysis is based on the LCPI reformulation \cite{AZ} of the AMY 
formalism \cite{AMY1} to the photon emission from relativistic plasmas. 
The developed approach, contrary to the previous qualitative studies 
\cite{Gale,Zh1,Harko2},
allows, for the first time, to give a robust treatment
of the Pauli-blocking effects in the photon bremsstrahlung.
We demonstrate that for the temperatures $T\sim 0.1\div 1$ MeV
the dominating contribution to the photon
emission is due to bending of electron trajectories 
in the mean electric field of the electrosphere.
The energy flux from the mean field photon emission
is of order of the black body limit.
Our results show that the contribution 
of the Bethe-Heitler bremsstrahlung due to electron-electron interaction
is negligible as compared to the mean field photon emission.
In contrast with  \cite{Gale} we demonstrate that the LPM 
suppression is negligible.

The energy flux related to the mean field bremsstrahlung
turns out to be larger than that from the tunnel $e^{+}e^{-}$ pair creation
\cite{Usov1,Usov2} as well.
In the light of these results the situation with distinguishing 
bare quark stars made of the SQM in normal (or 2SC) phase 
from neutron stars may be more optimistic
than in the scenario with the tunnel $e^{+}e^{-}$ creation discussed
in \cite{Usov-LC}. 

\vspace {.7 cm}
\noindent
{\large\bf Acknowledgements}

\noindent
I would like to thank J.F. Caron for providing the file 
for the radiated energy flux obtained in \cite{Zh1}. I am also grateful
to T. Harko and D. Page for communication. 
This work is supported in part by the grant SS-6501.2010.2.

\vskip .5 true cm
\renewcommand{\theequation}{A.\arabic{equation}}
\setcounter{equation}{0}
\noindent  {\Large \bf Appendix A. 
Calculation of the function $D(\qb_{\perp})$}
\vskip .2 true cm 

In this appendix, we discuss  the calculation of the 
function $D(\qb_{\perp})$.
To evaluate this function one needs to know
the correlator $D^{\mu\nu}$.
In momentum representation one can obtain 
$$
D^{\mu\nu}(q)=-2[1+n_{B}(q_{0})]
\mbox{Im}D^{\mu\nu}_{r}(q)\,,
$$ 
where $n_{B}=[\exp(q_{0}/T)-1]^{-1}$ is the
Bose-Einstein factor, and
$D^{\mu\nu}_{r}(q)$ 
retarded Green's function. As was already noted the function $P(\rho)$ is 
gauge invariant, and
one can use $D^{\mu\nu}_{r}$ in any gauge.
Expressing the retarded propagator in the Coulomb gauge
in terms   
terms of longitudinal and transverse  
photon self-energies  one can obtain
\bea
D(\qbt)=-\frac{1}{\pi}\int\limits_{-\infty}^{\infty}
dq_{0} 
\frac{\exp(q_{0}/T)}{\exp(q_{0}/T)-1}
\left\{
\frac{{\rm Im}\Pi_{L}(q_{0},\qb)}{[\qb^{2}-{\rm Re}\Pi_{L}(q_{0},\qb)]^{2}
+({\rm Im}\Pi_{L}(q_{0},\qb))^{2}}
\right.\nonumber\\
\left.
+
\left.\frac{\qbt^{2}}{\qb^{2}}\dot
\frac{{\rm Im}\Pi_{T}(q_{0},\qb)}{[\qbt^{2}+{\rm Re}\Pi_{T}(q_{0},\qb)]^{2}
+({\rm Im}\Pi_{T}(q_{0},\qb))^{2}}
\right\}\right|_{q_{z}=q_{0}}\,\,.
\label{eq:70}
\eea
In numerical calculations we use for $\Pi_{L,T}$ the 
HDL expressions \cite{HDL1,HDL2}
\beq
\Pi_{L}(q_{0},\qb)=m_{D}^{2}\left[\frac{q_{0}}{2q}\ln\left(
\frac{q_{0}+q}{q_{0}-q}\right)-1\right]\,,
\label{eq:80}
\eeq
\beq
\Pi_{T}(q_{0},\qb)=\frac{m_{D}^{2}}{2}\left[
\frac{q_{0}^{2}}{q^{2}}+
\frac{(q^{2}-q_{0}^{2})q_{0}}{2q^{3}}
\ln\left(
\frac{q_{0}+q}{q_{0}-q}\right)-1\right]
\label{eq:90}
\eeq
with the Debye mass 
$m_{D}^{2}=\frac{4\alpha}{\pi}(\mu^{2}+\pi^{2}T^{2}/3)$.

\vskip .5 true cm
\renewcommand{\theequation}{B.\arabic{equation}}
\setcounter{equation}{0}
\noindent  {\Large \bf Appendix B.
Formulas for the light-cone wave functions}
\vskip .1 true cm 

For zero $\fb$ the light-cone wave functions 
have definite orbital quantum number $m$.
As was mentioned the light-cone wave functions appear from
the longitudinal integrals of the Green's function.
For $\fb=0$ it is the free Green's function given by
\beq
{\cal K}_{v}(\xi_2,\r_{2}|\xi_1,\r_{1})
\!=\!\frac{M}{2\pi i\xi}
\exp
\left\{i\left[
\frac{M(\r_{2}-\r_{1})^{2}}{2 \xi}
-\frac{\xi\epsilon^{2}}{2M}
\right]
\right\}
\label{eq:a1}
\eeq
with $\xi=\xi_{2}-\xi_{1}$.
The $\xi$-integration can be performed with the help of the relation
\beq
\int_{-\infty}^{0} d\xi {\cal K}_{v}(\r_{2},0|\r_{1},\xi)=
-\frac{iM}{\pi}K_{0}(|\r_{2}-\r_{1}| \epsilon)\,,
\label{eq:a2}
\eeq
where $K_{0}$ is the Bessel function.
Then the light-cone wave functions can be written
in terms of the Bessel functions $K_{0}$ and $K_{1}$.
After representing the vertex operator (\ref{eq:31})
in terms of the helicity state projectors 
as in \cite{AZ} one can obtain
for $\lambda_{e'}=\lambda_{e}$
\beq
\Psi(x,\r,\lambda_{e},\lambda_{e'},\lambda_{\gamma})=
\frac{1}{2\pi}\sqrt{\frac{\alpha
}{2x}}
\left[\lambda_{\gamma}(2-x)+2\lambda_{e}x\right]
\exp(-i\lambda_{\gamma}\varphi)\epsilon K_{1}(\rho
\epsilon)\,,
\label{eq:128}
\eeq
where $\varphi$ is the azimuthal angle. For 
$\lambda_{e'}=-\lambda_{e}$ 
\beq
\Psi(x,\r,\lambda_{e},-\lambda_{e},2\lambda_{e})=
\frac{-i}{2\pi}\sqrt{2\alpha x^{3}}m_{e}K_{0}(\rho
\epsilon)\,.
\label{eq:130}
\eeq

\newpage

\begin{figure}[htb true]
\begin{center}
\epsfig{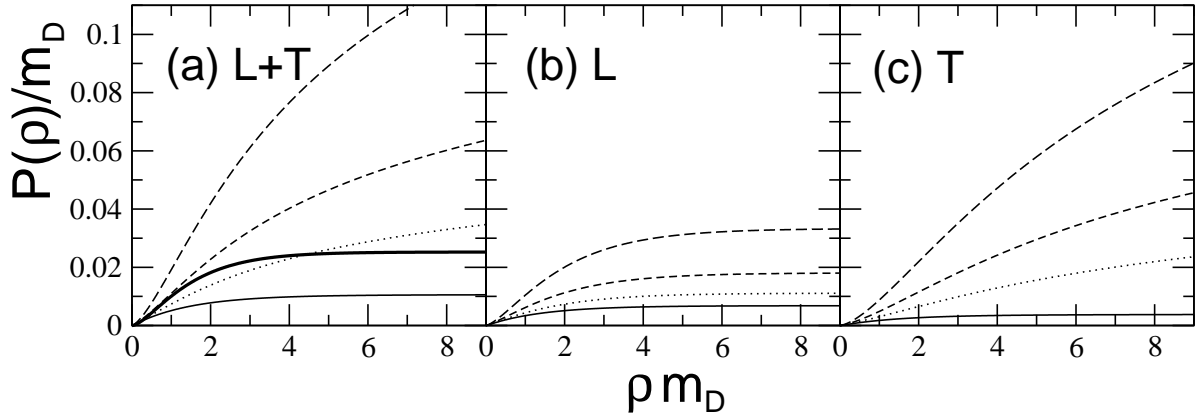}
\end{center}
\caption[.]{
The function $P(\rho)$ (\ref{eq:60}) in units of the Debye mass versus 
$\rho m_{D}$
for different values of the ratio $\tau=T/m_{D}$.
(a) shows the total $L+T$ contribution, (b) and (c) show $L$ and $T$
contributions, respectively. 
The curves correspond to: $\tau=0$ (solid line), $\tau=0.5$ (dotted line),
$\tau=1$ (short dashes), $\tau=2$ (long dashes).
The thick solid line in panel (a) shows prediction of the static model
obtained with the dipole cross section (\ref{eq:91}).
}
\end{figure}
\begin{figure}[htb]
\begin{center}
\epsfig{file=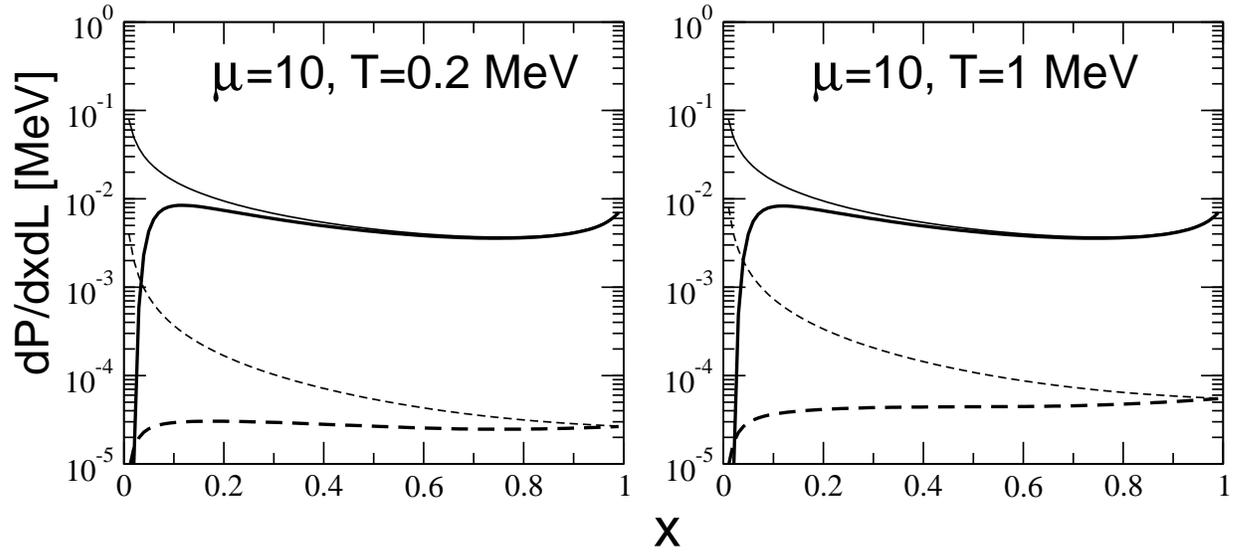,height=14.7cm}
\end{center}
\caption[.]{
The contributions to the spectrum $dP/dxdL$ 
from the mean field mechanism (solid line) and the fluctuation mechanism
(dashes) for $\mu=10$ MeV at $T=0.2$ and $T=1$ MeV. The thick curves are
for nonzero photon mass, and the thin ones are for massless photon.
The contribution of the fluctuation mechanism is calculated using 
the Bethe-Heitler term with the distribution  (\ref{eq:140}).
}
\end{figure}
\begin{figure}[htb]
\begin{center}
\epsfig{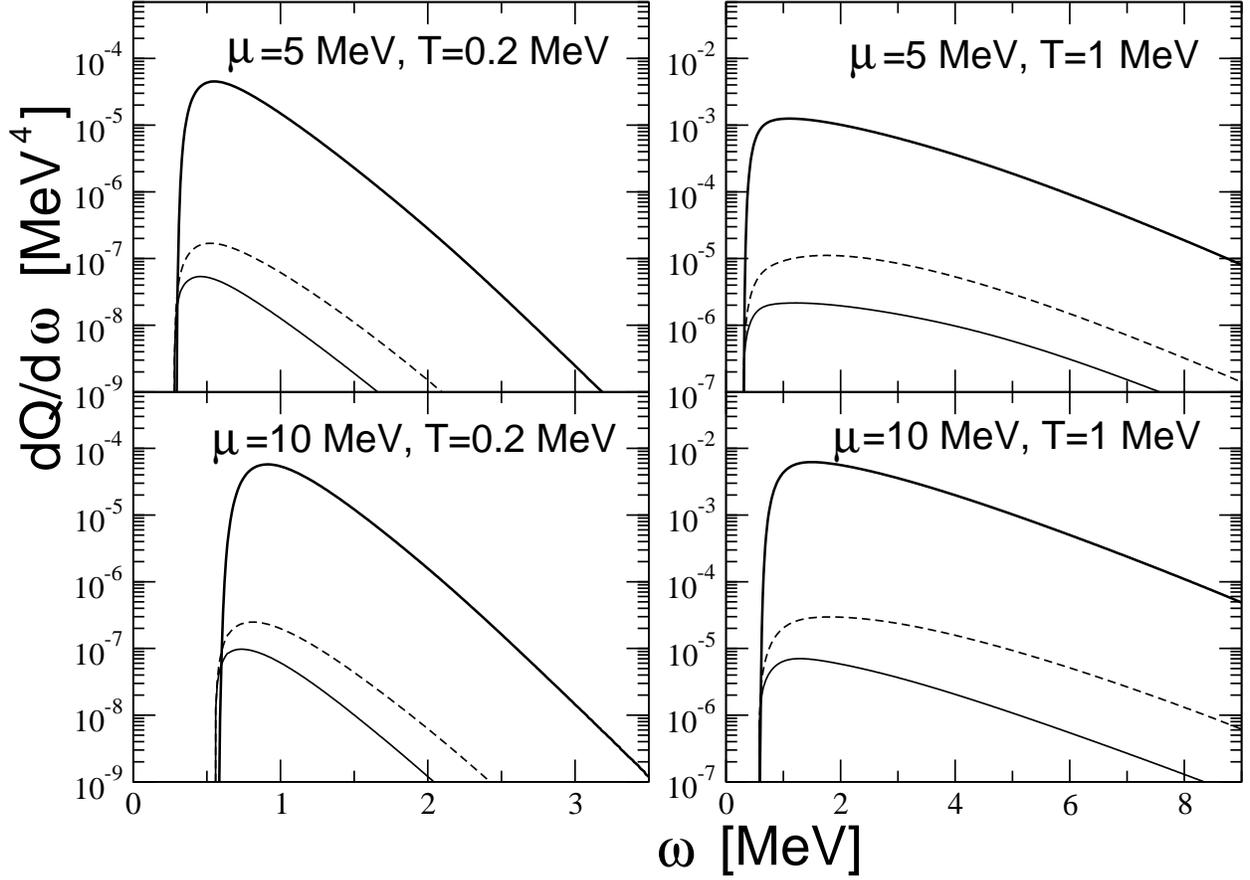}
\end{center}
\caption[.]{
The emissitivity versus the photon energy $\omega$
for $\mu=5$ and $\mu=10$ MeV at $T=0.2$ and $T=1$ MeV.
The thick solid line shows the mean field
bremsstrahlung. The contribution of the fluctuation 
mechanism is shown 
without (dashes) and with (thin solid line) the mean field suppression.
}
\end{figure}
\begin{figure}[htb]
\begin{center}
\epsfig{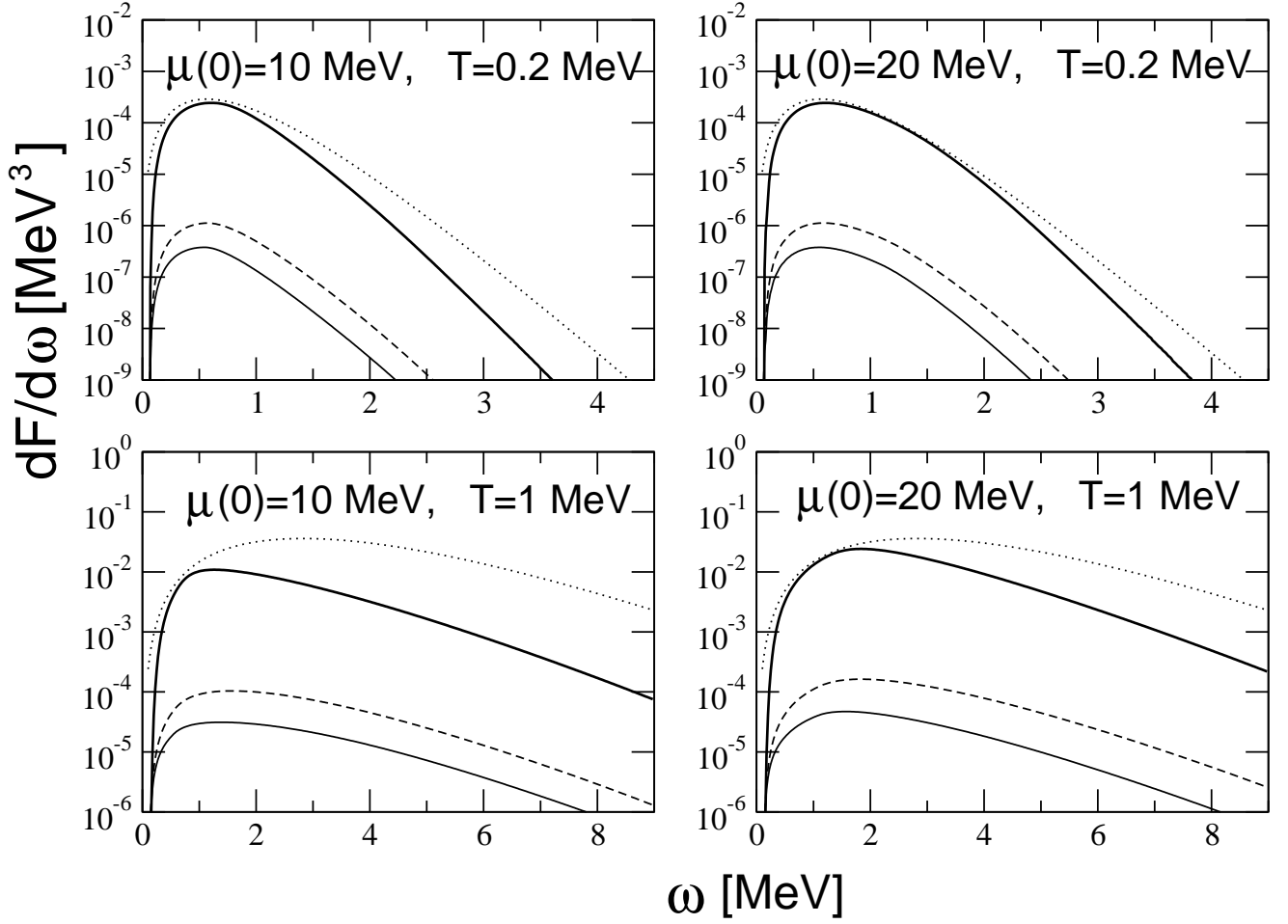}
\end{center}
\caption[.]{
The differential radiated energy flux from the electrosphere
for the mean field
bremsstrahlung (thick solid line) and for the Bethe-Heitler bremsstrahlung 
with (thin solid line) and 
without (dashes) the mean field suppression.
The dotted curves show the black body
spectrum.
}
\end{figure}
\begin{figure}[htb] 
\begin{center}
\epsfig{file=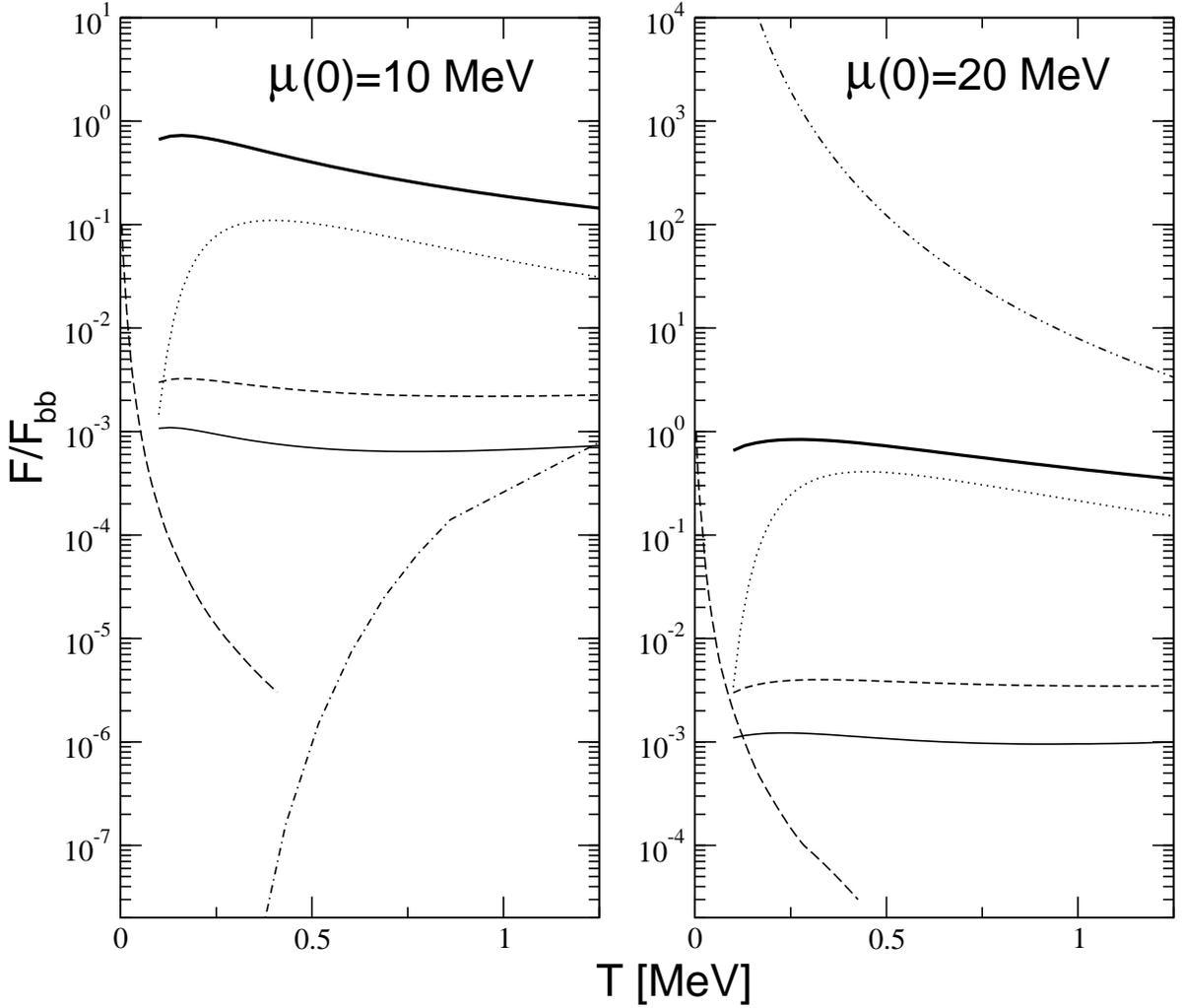,height=14.3cm}
\end{center}
\caption[.]{
The total radiated energy flux 
(scaled to the black body radiation)
from the electrosphere
for the mean field
bremsstrahlung (thick solid line) and for the Bethe-Heitler bremsstrahlung
with (thin solid line) and 
without (short dashes) the mean field suppression.
The contribution from the 
the tunnel $e^{+}e^{-}$ creation \cite{Usov1,Usov2}
evaluated using (\ref{eq:170}) is also shown (dotted line).
The long dashes show the results for
$e^{-}+e^{-}\to e^{-}+e^{-}+\gamma$ process obtained in
\cite{Gale}.
The dot-dashed line show the results for 
the same process of \cite{Zh1}. 
The dot-dot-dashed line shows the bremsstrahlung contribution
with inclusion of the mean Coulomb field of \cite{Harko2}.
}
\end{figure}

\end{document}